\documentclass[aps,prb,reprint,groupedaddress]{revtex4-1}
\usepackage{graphicx}
\usepackage{amsmath}
\usepackage{subfigure}
\usepackage{color}
\usepackage{float}
\usepackage{longtable}
\newcommand{\MI}[1]{\color{black}{#1} \color{black}}
\newcommand{\MIp}[1]{\color{black}{#1} \color{black}}
\newcommand{\MIPRB}[1]{\color{black}{#1} \color{black}}
\begin{document}


\title{Symmetric and antisymmetric strain as continuous tuning parameters for electronic nematic order}



\author{M.\,S.\,Ikeda}
\author{T.\,Worasaran}
\author{J.\,C.\,Palmstrom}
\author{J.\,A.\,W.\,Straquadine}
\author{P.\,Walmsley}
\author{I.\,R.\,Fisher}

\affiliation{Geballe Laboratory for Advanced Materials and Department of Applied Physics, Stanford University, Stanford, CA 94305, USA}
\affiliation{Stanford Institute for Materials and Energy Science, SLAC National Accelerator Laboratory, 2575 Sand Hill Road,
Menlo Park, California 94025, USA}

\date{\today}

\begin{abstract}
We report the separate response of the critical temperature of the nematic phase transition $T_{\rm S}$ to symmetric and antisymmetric
strains for the prototypical underdoped iron pnictide Ba(Fe$_{0.975}$Co$_{0.025}$)$_2$As$_2$. This decomposition is achieved by comparing the response of $T_{\rm S}$ to in-plane uniaxial stress and hydrostatic pressure. In addition to quantifying the two distinct linear responses to symmetric strains, we find a quadratic variation of $T_{\rm S}$ as a response to antisymmetric strains $\MI{\varepsilon}_{\rm B_{1g}}$=$\frac{1}{2}$($\MI{\varepsilon}_{\rm xx}$-$\MI{\varepsilon}_{\rm yy}$), exceeding the non linear response to symmetric strains by at least two orders of magnitude. These observations establish orthogonal antisymmetric strain as a powerful tuning parameter for nematic order.
\end{abstract}


\maketitle

\section{Introduction}
Electronic nematic order is found in several families of Fe-based superconductors \cite{Fer12.1,Fer14.1,Chua10.1,Chu10.1,Chu12.1,Boe14.1,Hos16.1} and also suggested to be an important aspect within the phase diagram of at least some cuprate high-temperature superconductors \cite{Ach16.1, Wu17.1,And02.1,Hin08.1,Dao10.1,Cyr15.1,Fuj14.1}. In order to assess the relevance of nematic fluctuations for superconductivity \cite{Kiv98.1,Bae04.1,Mai14.1,Met15.1,Led15.1,Kuo16.1,Led17.1,Nie17.1}, new methods are required to continuously tune the critical temperature of the nematic phase transition, with the ultimate goal of potentially providing access to a nematic quantum phase transition with a smoothly adjustable external parameter. Here we show how symmetric and antisymmetric strains induced by external stresses can be used as separate tuning parameters for nematic order. We demonstrate this for a representative underdoped Fe-pnictide, Ba(Fe$_{0.975}$Co$_{0.025}$)$_2$As$_2$, but emphasize that these ideas are quite general for nematic order. More broadly, the notions of symmetry decomposition that we employ can be applied to \MI{access strains as tuning parameters} for other types of phase transitions, \MI{thus offering a road map to gain further insight into most existing stress based phase diagrams.}

\begin{figure}[H]
	\centering
		\includegraphics[width=0.47\textwidth]{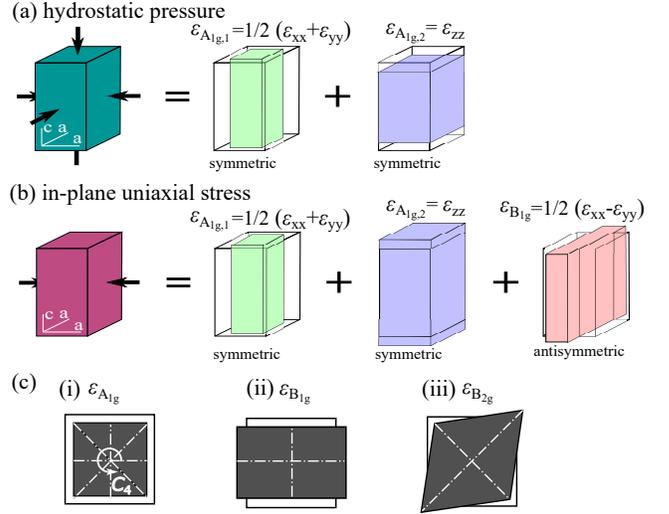}
	\caption{Schematic representation of strains experienced by a tetragonal material while held under (a) hydrostatic pressure, and (b) uniaxial stress applied along the [100] direction. Black arrows indicate stress. \MI{The strain tensor (right side of symbolic equations (a) and (b)) is derived by multiplying the stiffness and the stress tensor (left side).} White arrows in panel (a) and (b) indicate the orientation of the tetragonal crystal axes. In each case, \MI{the strain tensor} experienced by the material is decomposed into irreducible representations of the crystal symmetry. For materials with regular mechanical properties (i.e. a positive out-of-plane Poisson ratio and an in-plane Poisson ratio smaller than 1), the symmetric strain modes $\MI{\varepsilon}_{\rm A_{1g},i}$ share the same sign (are both compressive) during hydrostatic pressure experiments, but have opposite sign during uniaxial stress experiments. Panel (c) illustrates in-plane deformations as well as the associated preserved symmetries (white lines). While symmetric $A_{\rm 1g}$ strain preserves $C_4$ rotational symmetry (white arrow) as well as vertical, horizontal, and diagonal mirror planes (white dash dotted lines), antisymmetric $B_{\rm 1g}$ and $B_{\rm 2g}$ strains lower the primary rotational symmetry to $C_2$ and break diagonal and vertical mirror planes, respectively.} 
	\label{fig:sketch}
\end{figure}

The irreducible representations of the crystallographic point group provide a natural basis in which to express strains experienced by a solid. Within the D$_{\rm 4h}$ point group, appropriate for the specific material discussed in this paper, the 6 independent components of the strain tensor can be decomposed into two components that are symmetric with respect to the primary ($C_4$) rotation of the point group ($\MI{\varepsilon}_{\rm A_{1g},1}=\frac{1}{2}\left(\MI{\varepsilon}_{\rm xx}+\MI{\varepsilon}_{\rm yy}\right)$, $\MI{\varepsilon}_{\rm A_{1g},2}=\MI{\varepsilon}_{\rm zz}$, \MI{Fig.\,\ref{fig:sketch}(c)(i))}, two components that are antisymmetric ($\MI{\varepsilon}_{\rm B_{1g}}=\frac{1}{2}\left(\MI{\varepsilon}_{\rm xx}-\MI{\varepsilon}_{\rm yy}\right)$ and $\MI{\varepsilon}_{\rm B_{2g}}=\MI{\varepsilon}_{\rm xy}$, \MI{Fig.\,\ref{fig:sketch}(c)(ii) and (iii))} and two components that belong to an $E_{\rm g}$ representation, comprising vertical shear, $\MI{\varepsilon}_{\rm E_g}=\left(\MI{\varepsilon}_{\rm xz},\MI{\varepsilon}_{\rm yz}\right)$. The specific challenge is to separately determine the effects of each of these strain components on the critical temperature \MI{$T_{\rm c}$} of a phase transition.
In this paper we demonstrate how this can be achieved by comparison of the response to two (or more) different stress conditions (illustrated in Fig.\,\ref{fig:sketch}(a,b)). This decomposition not only reveals the relative roles of in-plane and out-of-plane symmetric strains, but also establishes orthogonal antisymmetric strain as a powerful new tuning parameter for nematic order.

\MI{As broken symmetries are the most fundamental organizing principle for (solids and) phase transitions, we start by reviewing the symmetry constraints for the nematic transition in underdoped Fe-pnictide superconductors, like Ba(Fe$_{0.975}$Co$_{0.025}$)$_2$As$_2$ studied here}. The nematic order parameter and associated lattice distortion that onset at the coupled nematic/structural phase transition at $T_{\rm S}$ have a $B_{\rm 2g}$ symmetry \MI{(broken $C_4$ rotation and horizontal and vertical mirrors/rotations)}. Hence, an external stress that induces an antisymmetric strain with a $B_{\rm 2g}$ symmetry (Fig.\,\ref{fig:sketch}(c)(iii)) necessarily induces a finite order parameter at all temperatures and therefore smears the phase transition \cite{Chu10.1,Dhi12.1,Lu16.1,Kis18.1}. Stresses that induce the \emph{orthogonal} antisymmetric strain, $\MI{\varepsilon}_{\rm B_{1g}}$, however, preserve the horizontal and vertical mirrors/rotations (Fig.\,\ref{fig:sketch}(c)(ii)). Consequently, a nematic phase transition is still permitted and $\MI{\varepsilon}_{\rm B_{1g}}$ can therefore be used as a continuous tuning parameter for the phase transition. Since $T_{\rm S}$ is an isotropic quantity \MI{(invariant under ($C_4$) rotation)}, antisymmetric strain $\MI{\varepsilon_{\rm B_{1g}}}$ can affect $T_{\rm S}$ only in even powers\MI{, thus $\lambda_{\rm B_{1g}}\equiv\partial T_{\rm S}/\partial\varepsilon_{\rm B_{1g}}=0$ ($\lambda_{\rm (B_{1g},A_{1g},i)}\equiv\partial^2 T_{\rm S}/\partial\varepsilon_{\rm B_{1g}}\partial\varepsilon_{\rm A_{1g},i}$=0)}. In contrast, the two symmetric strain components $\MI{\varepsilon}_{\rm A_{1g},1}$ and $\MI{\varepsilon}_{\rm A_{1g},2}$ do not lower the crystal symmetry (Fig.\,\ref{fig:sketch}(c)(i)) and therefore to leading order affect $T_{\rm S}$ linearly. Hence, considering both $A_{\rm 1g}$ and $B_{\rm 1g}$ symmetry strains, to second order the critical temperature $T_{\rm S}$ is given by
\begin{equation}
\begin{split}
&T_{\rm S}=T_{\rm S}(0)  + \sum_{i=1}^2\MI{\lambda}_{\rm (A_{1g,i})} \MI{\varepsilon}_{A_{1g,i}} +\\
&+\sum_{i\leq j=1}^2\MI{\lambda}_{(A_{1g,i},A_{1g,j})} \MI{\varepsilon}_{A_{1g,i}}\MI{\varepsilon}_{A_{1g,j}} + \MI{\lambda}_{(B_{1g},B_{1g})} \MI{\varepsilon}_{B_{1g}}^2,
\end{split}
\end{equation}
defining the coefficients $\MI{\lambda}_{\rm i}$ that are to be measured. This is achieved by comparing measurements of $T_{\rm S}(\MI{\varepsilon})$ obtained from uniaxial stress and hydrostatic pressure experiments.

\section{Experimental Methods}
\begin{figure}[ht!]
\centering	
\includegraphics[width=0.50\textwidth]{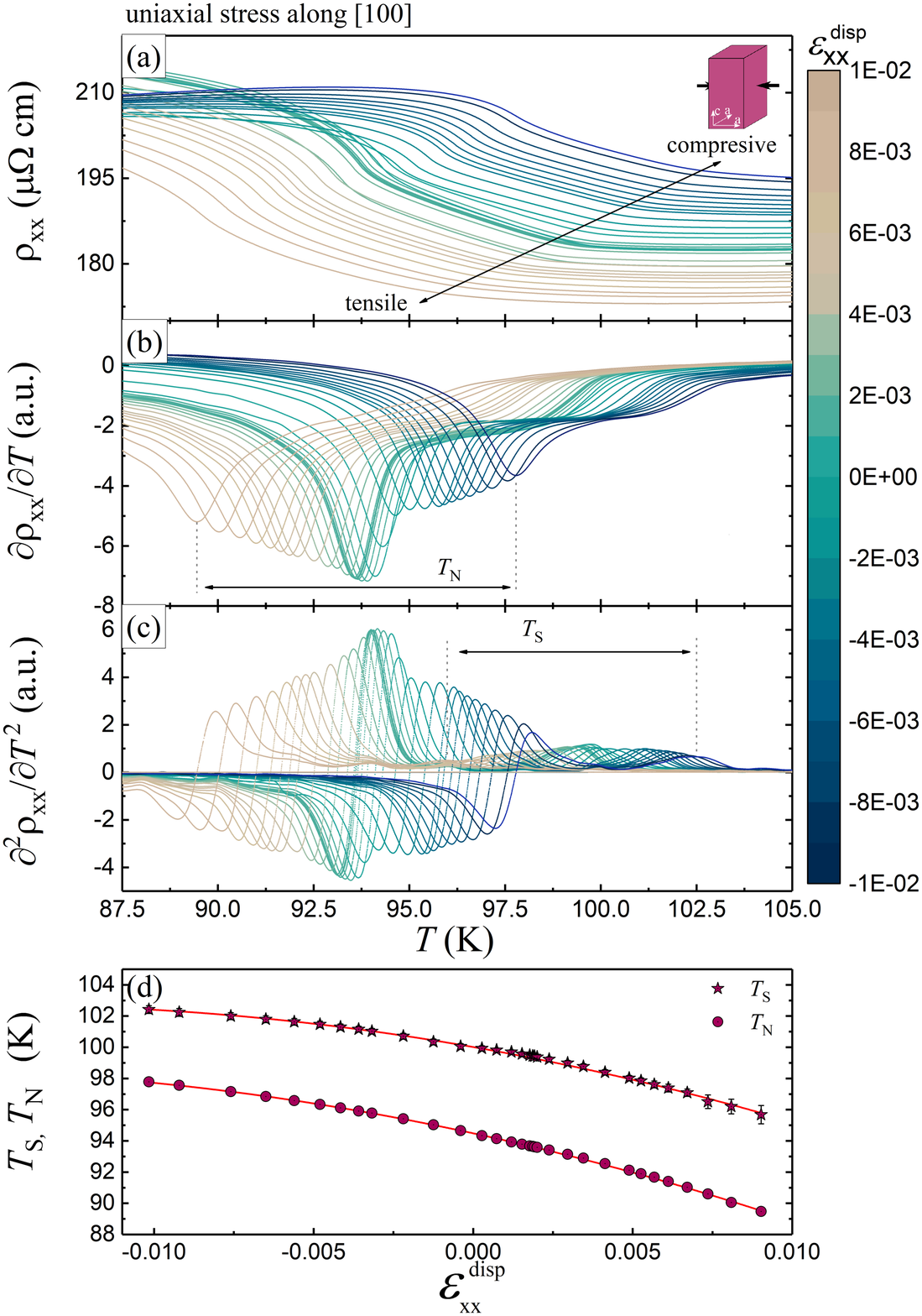}
\caption{(a) Electrical resistivity of Ba(Fe$_{0.975}$Co$_{0.025}$)$_2$As$_2$ as a function of temperature determined during a uniaxial stress experiment. For each temperature sweep (warming) shown here, the sample is held at a constant strain, $\MI{\varepsilon}_{\rm xx}^{\rm disp}$, indicated by the color scale (blue data points indicate compressive strain, beige points tensile strain, and cyan colored data small strain around the strain neutral point). Panel (b) and (c) show the first and second derivative of the resistivity with respect to temperature. Panel (d) shows $T_{\rm S}$ and $T_{\rm N}$ versus nominal strain $\MI{\varepsilon}_{\rm xx}^{\rm disp}$. The red lines are fits using a second order polynomial resulting $T_{\rm S}(\MI{\varepsilon}_{\rm xx})=100- (521\pm4) \MI{\varepsilon}_{\rm xx}-(28300\pm1100)\MI{\varepsilon}_{\rm xx}^2$, and $T_{\rm N}(\MI{\varepsilon}_{\rm xx})=94.5-(629\pm2) \MI{\varepsilon}_{\rm xx}-(24700\pm500)\MI{\varepsilon}_{\rm xx}^2$.}
\label{fig:RZBsummary}
\end{figure}

\begin{figure}[ht!]
	\centering
		\includegraphics[width=0.50\textwidth]{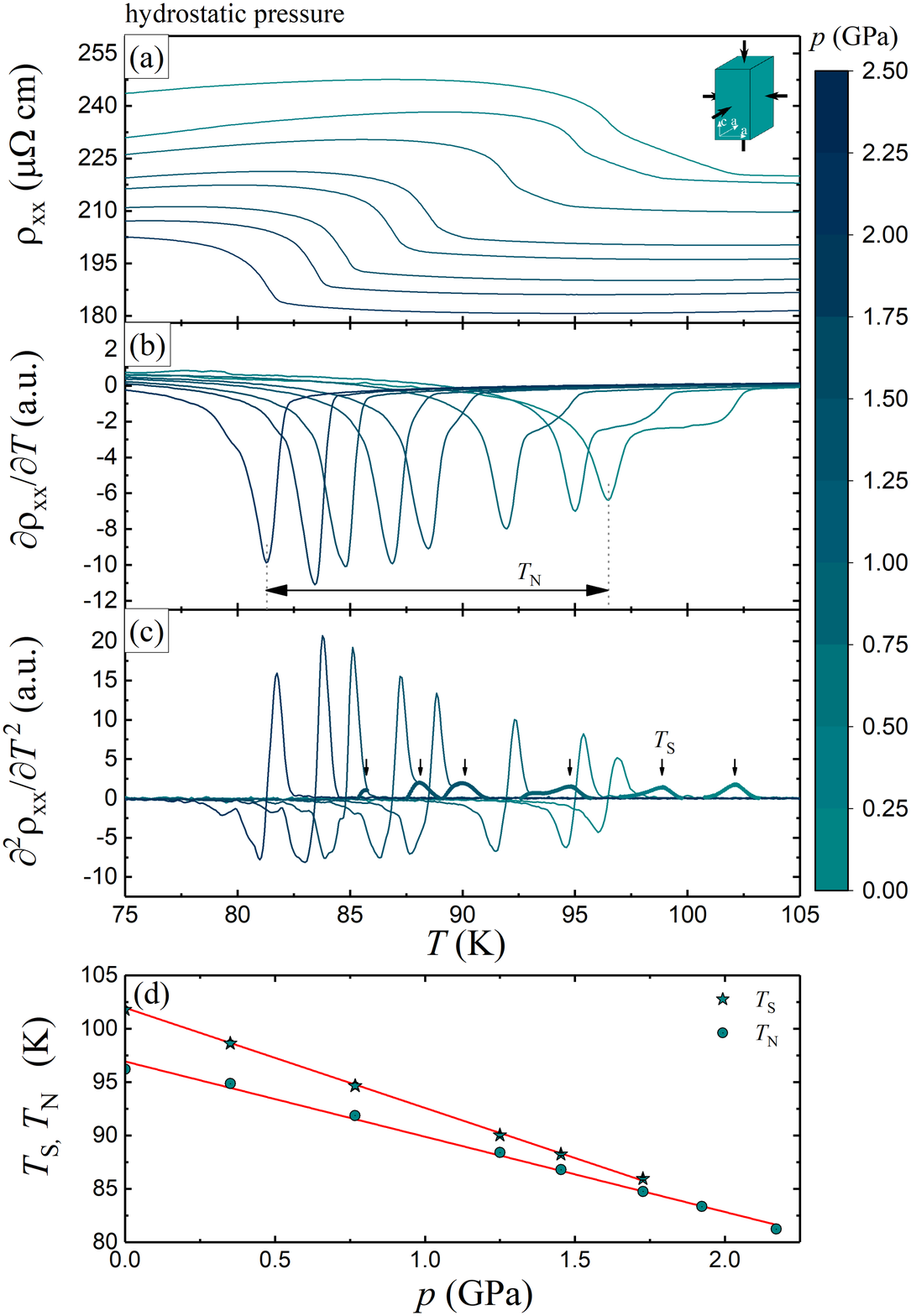}
		\caption{(a) Temperature dependence of the electrical resistivity $\rho$ of Ba(Fe$_{0.975}$Co$_{0.025}$)$_2$As$_2$ for a range of hydrostatic pressures $p$ determined during slow warming temperature sweeps. Panel (b) and (c) show the first and second derivative with respect to temperature, respectively. The dotted lines in panel (c) shows second derivatives calculated neglecting resistivity data below $T_{\rm N}$ +1K, \MI{avoiding overlap of the smoothed signatures of the features associated with $T_{\rm N}$ and $T_{\rm S}$}. Similar to Fig.\,\ref{fig:RZBsummary}, blue colored symbols indicate compressive strains. Panel (d) again shows the structural and antiferromagnetic transition temperature as a function of pressure. The red lines in panel (d) are linear fits to the data.}
	\label{fig:HPCsummary}
\end{figure}

\MIp{For both hydrostatic, and uniaxial stress experiments, the transition temperatures are determined from resistivity data (Fig.\,\ref{fig:RZBsummary},\ref{fig:HPCsummary} (a)). The longitudinal resistivity $\rho_{\rm xx}$ as a function of temperature was determined during slow temperature sweeps (down and up for each $\MI{\varepsilon}_{xx}$) using a standard four probe technique \MIPRB{(see Appendix\,\ref{sec:Exp})} \MI{on a crystal contacted using PbSn reflow solder (\MIPRB{for more details see Appendix\,\ref{sec:Sampleprep}}).  The coupled structural/nematic transition temperature $T_{\rm S}$ was determined from the center of a Gaussian function fit to a local maximum in the second derivative (Fig.\,\ref{fig:RZBsummary},\ref{fig:HPCsummary} (c)), the magnetic transition temperature $T_{\rm N}$ from a minimum in the first derivative (Fig.\,\ref{fig:RZBsummary},\ref{fig:HPCsummary} (b)). An upper bound for the error bar around $T_{\rm S}$ is estimated by half the standard deviation of the Gaussian function \footnote{The 95\% confidence interval of the center of the Gaussian function is smaller than $\pm$50\,mK in each case. To also take into account systematic uncertainty introduced by data smoothing or the potentially asymmetric shape of the peak around $T_{\rm S}$, we empirically determined half the standard deviation as an upper bound for the experimental uncertainty.}.}}

Uniaxial stress experiments were performed using a commercially available strain apparatus (CS100, \textit{Razorbill instruments}). Uniaxial stress was applied along a bar shaped sample of Ba(Fe$_{0.975}$Co$_{0.025}$)$_2$As$_2$ (with typical dimension 2000x400x35$\mu$m) by affixing it in between two mounting plates that are pushed together/pulled apart using voltage controlled lead zirconate titanate (PZT) stacks.  The cell is designed to compensate for the thermal expansion of the PZT stacks \cite{Hic14.1}. Furthermore, due to matching of the thermal expansion of Ba(Fe$_{1-x}$Co$_x$)$_2$As$_2$ \cite{Dal09.1} and the sample mounting plates (titanium) (\MIPRB{see Appendix\,\ref{sec:ExpUniax}, Fig.\,\ref{fig:thermalexpansion}}), the strain on the sample is almost perfectly independent of temperature for a fixed voltage applied to the PZT stacks. Stress was applied along the tetragonal [100] axis resulting in a combination of $\MI{\varepsilon}_{\rm A_{1g},1}$, $\MI{\varepsilon}_{\rm A_{1g},2}$, and $\MI{\varepsilon}_{\rm B_{1g}}$ (see Fig.\,\ref{fig:sketch} (b)). \MIPRB{The misalignment of the [100] crystal and the stress axis was estimated to be smaller than 2.5$^\circ$ by comparing to a uniaxial stress experiment on a crystal oriented such that stress was applied along the tetragonal [110] axis (see Appendix\,\ref{sec:ExpUniax110}).}
The nominal strain along the tetragonal [100] axis ($\MI{\varepsilon}_{\rm xx}^{\rm disp}$) was determined by the zero strain length of the sample between the mounting plates and the length change measured by sampling a capacitance sensor using an \textit{Andeen-Hagerling} AH2550A capacitance bridge. Due to strain relaxation effects in the mounting plates and the mounting glue, the actual strain $\MI{\varepsilon}_{\rm xx}$ experienced by the sample is smaller than $\MI{\varepsilon}_{\rm xx}^{\rm disp}$. Using finite element simulations (\MIPRB{see Appendix\,\ref{sec:Simul}}), we estimate $\MI{\varepsilon}_{\rm xx}=(0.7\pm0.07) \MI{\varepsilon}_{\rm xx}^{\rm disp}$. 
The extracted critical temperature can be well fit by $T_{\rm S}(\MI{\varepsilon}_{\rm xx})=T_{\rm S}(\MI{\varepsilon}_{\rm xx}=0)+\alpha\MI{\varepsilon}_{\rm xx}+\beta\MI{\varepsilon}_{\rm xx}^2$ (red line, \footnote{Note that Fig.\,\ref{fig:RZBsummary} (d) shows the fits on the original data, before correction for strain relaxation effects using $\MI{\varepsilon}_{\rm xx}=0.7 \MI{\varepsilon}_{\rm xx}^{\rm disp}$}), with $\alpha = -521 \pm 4$\,K and $\beta = -28300\pm 1100$\,K. As we will show later, the surprisingly large quadratic response is due solely to antisymmetric strain, $\MI{\varepsilon}_{\rm B_{1g}}$. 

The second experiment reported here is electrical resistivity on a bar shaped (1000x600x30$\mu$m) crystal of Ba(Fe$_{0.975}$Co$_{0.025}$)$_2$As$_2$ as a function of temperature under hydrostatic pressure (Fig.\,\ref{fig:HPCsummary}). The measurements were performed using a HPC-30 pressure cell within a PPMS from \textit{Quantum Design} \MI{using Daphne oil 7373 as pressure medium}. The hydrostatic pressure was determined by measuring the superconducting transition temperature of a lead manometer. Under perfectly hydrostatic conditions \MIPRB{(for details see Appendix\,\ref{sec:ExpHydro})} the strain is purely symmetric, and both $\MI{\varepsilon}_{\rm A_{1g},1}$ and $\MI{\varepsilon}_{\rm A_{1g},2}$ are compressive (Fig.\,\ref{fig:sketch}(a)).
Both transitions are found to vary almost perfectly linearly under hydrostatic pressure, though with a slightly different slope (see Fig.\,\ref{fig:HPCsummary}(d)), thus merging for pressures greater than approximately 1.75\,GPa. A linear fit results in $T_{\rm S}(P)$ = $T_{\rm S}(P=0) + \tilde\alpha P$, where $\tilde\alpha = -9.38\pm0.08\,{\rm K}/{\rm GPa}$.

\section{Results and Discussion}
\begin{figure}[ht]
\includegraphics[width=0.5\textwidth]{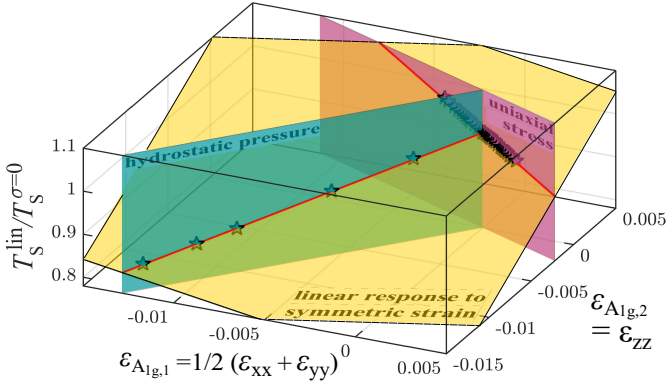}
\caption{Normalized linear response of the structural transition to symmetric $\MI{\varepsilon}_{\rm A_{1g},1}$, and $\MI{\varepsilon}_{\rm A_{1g},2}$ strain during hydrostatic pressure and uniaxial stress experiments. The cyan and purple vertical planes indicate the relative combination of symmetric strains induced during hydrostatic pressure and uniaxial stress experiments, respectively. Experimental data are shown by the cyan (hydrostatic pressure) and purple stars (uniaxial stress), respectively. Linear fits are shown by red lines. The yellow plane defined by these two lines describes the material's linear response to symmetric strain.}
\label{fig:Tsvseps}
\end{figure}

To decompose the strain induced changes of $T_{\rm S}$, the relative amount of strains within each of the different symmetry channels is first determined, based on the measured elastic stiffness tensor\cite{Yos12.1} (\MIPRB{see Appendix\,\ref{sec:SimulElast}}). The linear response (which, as described earlier, can only arise due to symmetric strain) can then be plotted as a function of the decomposed strains $\MI{\varepsilon}_{\rm A_{1g},1}$ and $\MI{\varepsilon}_{\rm A_{1g},2}$ (Fig.\,\ref{fig:Tsvseps}). For hydrostatic pressure (cyan plane in Fig.\,\ref{fig:Tsvseps}) the response is purely linear, so no subtraction is necessary. For uniaxial stress, the quadratic term ($\beta\MI{\varepsilon}_{\rm xx}^2$, defined earlier) is first subtracted using the fitted value of $\beta$, to leave the linear response: $T_{\rm S}^{\rm lin}(\MI{\varepsilon}_{\rm xx}) = T_{\rm S}(\MI{\varepsilon}_{\rm xx})-\beta \MI{\varepsilon}_{\rm xx}^2$ (purple stars in Fig.\,\ref{fig:Tsvseps}). Since the ratio of $\MI{\varepsilon}_{\rm A_{1g},1}$ and $\MI{\varepsilon}_{\rm A_{1g},2}$ is different for the two experiments, the purple and cyan planes in Fig.\,\ref{fig:Tsvseps} are not parallel. Moreover, since two lines define a plane, the material's response to symmetric strain (yellow plane in Fig.\,\ref{fig:Tsvseps}) is uniquely defined by these two sets of measurements. A full decomposition of the response to symmetric strains is now possible, and the associated partial derivatives are readily determined; $\MI{\lambda}_{\rm (A_{1g,1})}$=-6.35$\pm$0.23$\,{\rm K}/{\%}$, and $\MI{\lambda}_{\rm (A_{1g,2})}$=16.70$\pm$0.32$\,{\rm K}/{\%}$ \footnote{Error bars represent statistical uncertainty. A more comprehensive error analysis is given in Appendix\,\ref{sec:Decomp}} (for details see Appendix\,\ref{sec:Decomp}). The ratio of these terms $\MI{\lambda}_{\rm (A_{1g,2})}$/$\MI{\lambda}_{\rm (A_{1g,1})}$=-2.63$\pm$0.11 demonstrates that c-axis strains have a considerably larger effect on $T_{\rm S}$ than symmetric in-plane strains. Considering the contributions of symmetric strain to a standard Landau free energy expansion of the nematic phase transition (discussed further below), we note that the finite values of $\MI{\lambda}_{\rm (A_{1g,1})}$ and $\MI{\lambda}_{\rm (A_{1g,2})}$ necessarily imply the formation of spontaneous symmetric strains $\MI{\varepsilon}_{\rm A_{1g},1}$ and $\MI{\varepsilon}_{\rm A_{1g},2}$ upon cooling below $T_{\rm S}$. That $\MI{\lambda}_{\rm (A_{1g,1})}$ and  $\MI{\lambda}_{\rm (A_{1g,2})}$ have opposite signs, is consistent \footnote{The sign of $\MI{\varepsilon}_{\rm A_{1g},1}/\MI{\varepsilon}_{\rm A_{1g},2}$ is expected to match the sign of $\MI{\lambda}_{\rm (A_{1g,1})}$/$\MI{\lambda}_{\rm (A_{1g,2})}$ in case $C_{\rm A_{1g,1,1}},C_{\rm A_{1g,2,2}}, C_{\rm A_{1g,1,2}}>0$, where $C_{\rm A_{1g,i,j}}=\frac{\partial^2 F}{\partial\MI{\varepsilon}_{\rm A_{1g},i}\partial\MI{\varepsilon}_{\rm A_{1g},j}}$.} with the observation that the spontaneous strains as measured in recent thermal expansion experiments \cite{Mei12.1} have opposite signs.

\begin{figure}
\centering	
\includegraphics[width=0.5\textwidth]{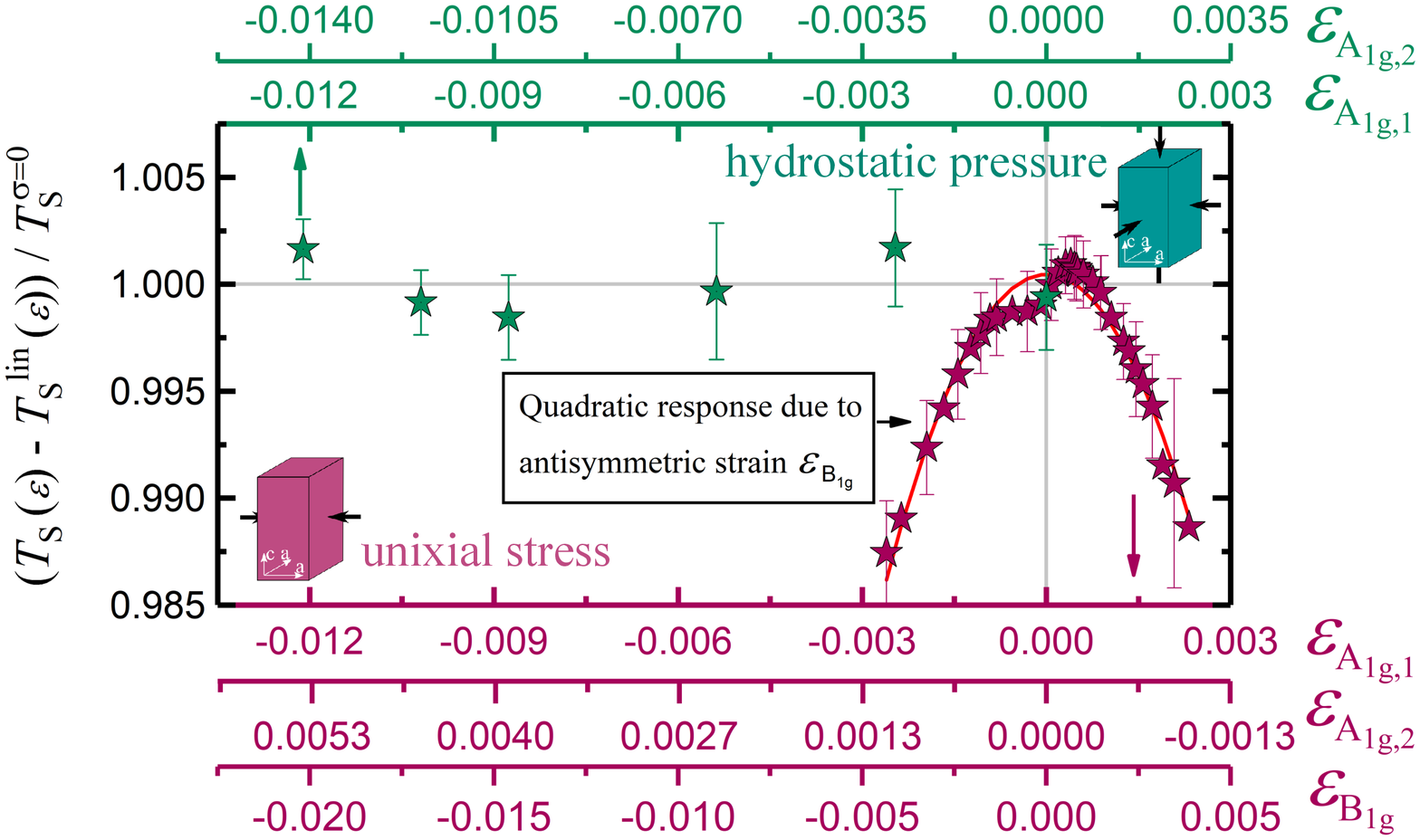}
\caption{Normalized non linear response of the coupled nematic/structural transition temperature to strain during hydrostatic (cyan stars, top axes) and uniaxial stress (purple stars, bottom axes) experiments. The quadratic contribution evident in the uniaxial stress data is solely due to antisymmetric $B_{1g}$ strain. Red line shows fit to a quadratic function. \MI{The origin of the kink near zero strain is currently unknown. The feature is, however, smaller than the error bars and has no statistically significant impact on the fit}.}
\label{fig:residuals}
\end{figure}

Having decomposed the linear response, we turn to the quadratic response. Figure\,\ref{fig:residuals} shows the normalized non-linear response of $T_{\rm S}$ to strain calculated by subtracting the linear response (i.e. subtracting $\alpha \MI{\varepsilon}_{\rm xx}$ and $\tilde\alpha P$ from the data shown in Figs.\,2(d) and 3(d) respectively, using the fitted values of $\alpha$ and $\tilde\alpha$). Data are shown as a function of the three (two) strain components present for the uniaxial stress (hydrostatic pressure) experiments on the bottom (top) axes. While quadratic responses to symmetric $A_{1g,i}$ strain are allowed by symmetry, no such response is observed during our hydrostatic pressure experiment (cyan data points). Moreover, $A_{1g}$ strains experienced in the hydrostatic pressure experiments exceed by nearly a factor of four those experienced by samples held under uniaxial stress. Hence, the quadratic response observed during our uniaxial stress experiment (purple curve in Fig.\,5) is caused exclusively by anti-symmetric $B_{1g}$ strain. The fit parameters then yield $\MI{\lambda}_{\rm (B_{1g},B_{1g})}=\partial^2T_{\rm S}/\partial\MI{\varepsilon}_{\rm B_{1g}}^2$=-7.25$\pm0.25\,{\rm K}/{\%^2}$. 

The quadratic functional form of $T_{\rm S}(\MI{\varepsilon}_{\rm B_{1g}})$ is understood based on symmetry, but the magnitude and sign of $\MI{\lambda}_{\rm (B_{1g},B_{1g})}$ are not determined by symmetry alone. Our measurements reveal that the ratio $\MI{\lambda}_{\rm (B_{1g},B_{1g})}$/$\MI{\lambda}_{\rm (A_{1g,i},A_{1g,i})}$ is at least 100, and possibly even larger \footnote{Quadratic fits to the hydrostatic pressure data yield $\MI{\lambda}_{\rm (A_{1g,1},A_{1g,1})}$=0.06$\pm0.07\,{\rm K}/{\%^2}$ and $\MI{\lambda}_{\rm (A_{1g,2},A_{1g,2})}$=0.05$\pm0.05\,{\rm K}/{\%^2}$. The quadratic response of $T_{\rm S}$ to $\MI{\varepsilon}_{\rm B_{1g}}$ is thus found to be at least two orders of magnitude larger. Since the uncertainty in $\MI{\lambda}_{\rm (A_{1g,i},A_{1g,i})} (i = 1,2)$ is so large, this is essentially a lower bound on the ratios $\MI{\lambda}_{\rm (B_{1g},B_{1g})}$/$\MI{\lambda}_{\rm (A_{1g,i},A_{1g,i})}$, which could therefore be considerably larger.}. \MIp{The surprisingly large value of $\MI{\lambda}_{\rm (B_{1g},B_{1g})}$ means that for tensile strains larger than $\MI{\varepsilon}_{\rm xx}^{\rm disp} \approx 1.84$\%, $\MI{\varepsilon}_{\rm B_{1g}}$} strain dominates the suppression of $T_{\rm S}$.
The physical origin of this very large effect remains to be determined, but we emphasize one important difference between strains of these two symmetries. Specifically, $A_{\rm 1g}$ strains do not break any symmetries (Fig.\,\ref{fig:sketch}c(i)) and therefore do not introduce any new terms to the low-energy effective Hamiltonian describing the system. However, $B_{1g}$ symmetry strain explicitly breaks specific symmetries (Fig.\,\ref{fig:sketch}c(ii)) and therefore introduces new operators to the effective Hamiltonian. Acting on an $E_{\rm g}$ orbital doublet (for example, degenerate $d_{\rm xz}$ and $d_{\rm yz}$ orbitals), operators with $B_{\rm 1g}$ and $B_{\rm 2g}$ symmetry do not commute; consequently $B_{\rm 1g}$ symmetry strain induces quantum fluctuations in a $B_{\rm 2g}$ symmetry order parameter \cite{Mah17.1}, possibly accounting for (or at least contributing to) the large negative value of $\MI{\lambda}_{\rm (B_{1g},B_{1g})}$.

On a phenomenological level, we note that the negative sign of $\MI{\lambda}_{\rm (B_{1g},B_{1g})}$ for Ba(Fe$_{0.975}$Co$_{0.025}$)$_2$As$_2$ is entirely consistent with observations that the $B_{1g}$ component of the elastic tensor, $\frac{1}{2}(c_{11}-c_{12})$ hardens upon cooling through $T_{\rm S}$ \cite{Yos12.1}. This can be readily appreciated by turning to the standard Landau treatment of the nematic phase transition \cite{Chu12.1}. Writing the free energy in even powers of the nematic order parameter and including coupling to strains $\MI{\varepsilon}_{\rm A_{1g},1}$, $\MI{\varepsilon}_{\rm A_{1g},2}$, $\MI{\varepsilon}_{\rm B_{1g}}$, and $\MI{\varepsilon}_{\rm B_{2g}}$ we obtain

\begin{equation}
\begin{split}
\Delta{F}=&\frac{a}{2}(T-T^*) \Phi_{\rm B_{2g}}^2 + \frac{b}{4} \Phi_{\rm B_{2g}}^4 + \MI{\lambda}_{B_{2g}} \MI{\varepsilon}_{\rm B_{2g}} \Phi_{\rm B_{2g}} - \\ 
        -&\MI{\frac{a}{2}}\MI{\lambda}_{\rm B_{1g}}\MI{\varepsilon}_{\rm B_{1g}}^2\Phi_{\rm B_{2g}}^2\MI{-\frac{a}{2}}\MI{\lambda}_{\rm A_{1g}}\MI{\varepsilon}_{\rm A_{1g}}^{\rm eff} \Phi_{\rm B_{2g}}^2+\frac{1}{2}C_{66}^{(0)}\MI{\varepsilon}_{\rm B_{2g}}^2 +\\
				+&\frac{1}{2}\left(\frac{C_{11}-C_{12}}{2}\right)^{(0)}\MI{\varepsilon}_{\rm B_{1g}}^2+\frac{1}{2}C_{\rm A_{1g}}^{\rm eff}\left(\MI{\varepsilon}_{\rm A_{1g}}^{\rm eff}\right)^2,\\
\end{split}
\end{equation}
where $\Phi_{\rm B_{2g}}$ is the nematic order parameter, the coupling coefficients $\MI{\lambda}_{\rm B_{1g}}$, $\MI{\lambda}_{\rm A_{1g,1}}$ and $\MI{\lambda}_{\rm A_{1g,2}}$ are determined by our measurements, and $\MI{\varepsilon}_{\rm A_{1g}}^{\rm eff}$ and $C_{\rm A_{1g}}^{\rm eff}$ are appropriate combinations of $A_{1g}$ symmetry strains and terms in the elastic stiffness tensor, determined by the Poisson ratio of the material. Significantly, the biquadratic coupling of $\Phi_{\rm B_{2g}}$ to $\MI{\varepsilon}_{\rm B_{1g}}$ not only provides means to tune $T_{\rm S}$, but also renormalizes the ‘bare’ elastic modulus $C_{\rm B_{1g}}=\left(\frac{C_{11}-C_{12}}{2}\right)^{(0)}$ for free-standing samples, such that 
\begin{equation}
C_{\rm B_{1g}}^{eff} = \frac{\partial^2F}{\partial\MI{\varepsilon}_{\rm B_{1g}}^2} = \left(\frac{C_{11}-C_{12}}{2}\right)^{(0)}\MI{-a\MI{\lambda}_{\rm B_{1g}}\Phi_{\rm B_{2g}}^2.}
\end{equation}
In other words, \MI{since $a>0$} our observation of \MI{a negative value of $\MI{\lambda}_{\rm B_{1g}}$=-7.25$\pm0.25\,\frac{\rm K}{\%^2}$} is consistent with observations that the $B_{1g}$ component of the elastic tensor, $\frac{C_{11}-C_{12}}{2}$ hardens \cite{Yos12.1} upon cooling through $T_{\rm S}$.

\section{Summary}

In-plane \MIp{strain} has previously been demonstrated as suitable means to induce phase transitions \cite{Boe17.1}. Here we have shown how a complete symmetry decomposition, made possible by comparison to hydrostatic pressure, reveals the separate effects of symmetric and antisymmetric strains that are necessarily both present when a sample is held under in-plane uniaxial stress. We emphasize that anti-symmetric strain is a powerful continuous tuning parameter for nematic phase transitions. While values of $\MI{\varepsilon}_{\rm B_{1g}}$ that would be necessary to completely suppress the coupled nematic/structural phase transition in Ba(Fe$_{0.975}$Co$_{0.025}$)$_2$As$_2$ are slightly out of reach, this is not necessarily the case for other materials, raising the distinct possibility that antisymmetric strain could be used to continuously tune a suitable material to a nematic quantum phase transition \cite{Mah17.1}. \MIPRB{Finally, we report that symmetric c-axis strain has a significantly stronger effect on the nematic transition in Ba(Fe$_{0.975}$Co$_{0.025}$)$_2$As$_2$ as compared to symmetric in-plane strain}.

\begin{acknowledgments}
The authors thank  J.\,Schmalian, S.\,A.\,Kivelson, A.\,Hristov, C.\,W.\,Hicks, and P.\,Massat for insightful discussions. This work was supported in part by the Gordon and Betty Moore Foundation’s EPiQS Initiative through Grant No. GBMF4414 (M.S.I \& P.W) and by the Department of Energy, Office of Basic Energy Sciences, under Contract No. DEAC02-76SF00515 (T.W \& I.R.F). J.C.P was supported by a NSF Graduate Research Fellowship (Grant No. DGE-114747) and a Gabilan Stanford Graduate Fellowship, J.A.W.S acknowledges support as an ABB Stanford Graduate Fellow.
\end{acknowledgments}

\appendix
\section{Sample preparation}
\label{sec:Sampleprep}
The Ba(Fe$_{0.975}$Co$_{0.025}$)$_2$As$_2$ single crystals characterized here were grown using a FeAs self flux technique described elsewhere \cite{Chu09.1}. The crystals were cleaved into thin plates and cut into rectilinear bars. Typical samples dimensions for uniaxial stress and hydrostatic pressure experiments were 2000$\times$400$\times$35$\mu$m and 1000$\times$300$\times$35$\mu$m, respectively. Electrical contacts were made by a reflow soldering technique using a Sn63Pb37 solder paste with a solder particle size of 15-25\,$\mu$m (Chip Quick SMD291AX10T5). The initial steps of the contacting procedure using solder paste are similar to contacting methods using silver paint or silver epoxy. The ends of short ($\sim$10\,mm) pieces of 25$\mu$m wide gold wire were dipped into the solder paste and positioned onto the freshly cleaved sample surface. The sample, resting on a 1 mm thick glass slide, is then placed on a hot plate preheated to 200\,$^\circ$C to reflow the solder beads. To prevent oxidation of the contacts, the sample and glass slide are taken off the hot plate as soon as the solder particles melt. This can be easily seen as the contact appearance changes from matte to shiny.
The typical contact resistance of such solder joints was estimated by a quasi 4 point measurement to be smaller than 20\,mOhm per contact. While the contact resistance of soldered contacts is superior to silver paint and silver epoxy, it is important to note that the superconductivity of the solder ($T_c\sim7.1$\,K) \cite{War69.1} might be problematic for measurements at low temperatures. These solder joints are also significantly more mechanically robust and are better able to survive thermal cycling than silver paint contacts.

\section{Experimental Methods}
\label{sec:Exp}
Four point resistivity measurements during our uniaxial stress and hydrostatic pressure experiments were performed using a Stanford Research Lock In amplifier (SR830). The output of the lock-in amplifier was converted to a constant current source using a 1\,kOhm series resistor. The voltage signal was amplified with a Stanford Research transformer preamplifier (SR554).  

All resistivity measurements used the same cryostat, a PPMS from Quantum Design. The temperature was swept slowly at a rate of 0.5K/min. The sample temperature was measured using a Cernox CX-1050 temperature sensor from Lakeshore mounted on the Ti body of the CS100 cell (for uniaxial stress experiments) and the Cu-Be body of the hydrostatic pressure cell, respectively. The temperature sensors were sampled using a Lakeshore 340 temperature controller. The thermal lag of the sample as compared to the measured cell temperature was estimated by taking resistivity measurements during cooling and warming runs. Thermal lags of about 0.1\,K and 0.25\,K were found for our uniaxial stress and hydrostatic pressure experiments, respectively.

\subsection{Uniaxial stress experiments}
\label{sec:ExpUniax}
\begin{figure}[ht]
	\centering
		\includegraphics[width=0.48\textwidth]{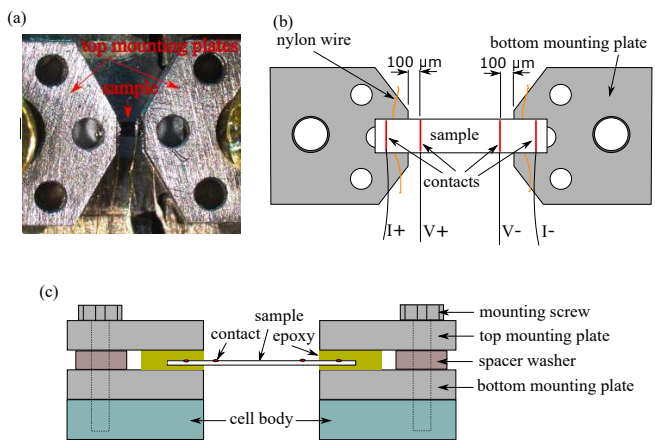}
	\caption{(a) Ba(Fe$_{0.92}$Co$_{0.08}$)$_2$As$_2$ crystal mounted on a Razorbill CS100 cell using bottom and top mounting plates. Panel (b) sketches a sample mounted onto bottom mounting plates, panel (c) sketches a cross sectional view.}
	\label{fig:sample}
\end{figure}

Uniaxial stress experiments were performed using a commercially available CS100 cell from Razorbill Instruments. This cell uses piezo electric (PZT) stacks to separate two mounting plates. The exact working principle of such a cell is described in detail elsewhere \cite{Hic14.1}. Samples were affixed onto the mounting plates of the uniaxial stress cell (see Figure\,\ref{fig:sample}) using Devcon 2-ton epoxy. The glue layer thickness between the sample and the bottom mounting plates was controlled using Nylon wire spacer with a diameter of 25$\mu$m. The glue layer thickness between the sample and the top mounting plates was controlled by the thickness of spacer washers between the top and bottom mounting plates. Typically, the glue layer on top of the sample was approximately double the thickness of the bottom glue layer.
    
\begin{figure}[ht]
	\centering
		\includegraphics[width=0.48\textwidth]{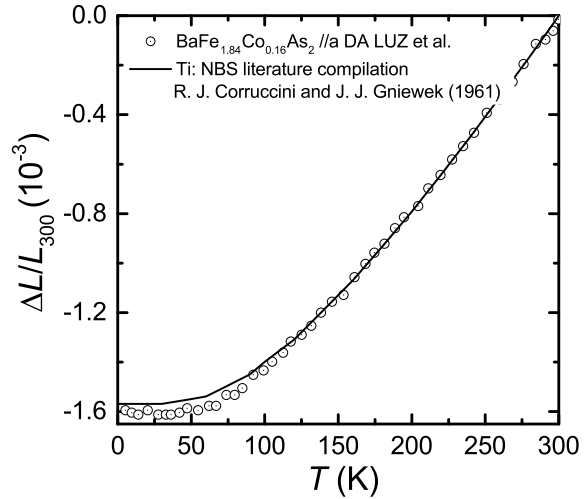}
	\caption{Comparison of the in-plane thermal expansion of Ba(Fe$_{0.92}$Co$_{0.08}$)$_2$As$_2$ \cite{Dal09.1} and the thermal expansion of Ti. \cite{Cor61.1}}
	\label{fig:thermalexpansion}
\end{figure}

The zero-volt strain (zero volts across the PZT stacks results in zero piezoelectric extension or contraction of the stacks) experienced by the sample during uniaxial stress experiments is determined by the differential thermal expansion of the sample and the sample mounting plates and thus, in general, is not temperature independent. In this case however, the thermal expansion of the Ti mounting plates and the in-plane expansion of Ba(Fe$_{0.92}$Co$_{0.08}$)$_2$As$_2$ are very similar (Fig.\,\ref{fig:thermalexpansion}). Therefore we can approximate the strain as independent of temperature and fully controlled by the voltages applied to the Piezo electric stacks. Due to the hysteresis of the PZT stacks, zero volts across all three PZT stacks does not necessarily correspond to zero strain.

The nominal strain along the transport direction ($\epsilon_{xx}^{disp}$) was determined by measuring the displacement of the two sides of the cell (using a capacitive displacement sensor sampled by an Andeen Hagerling AH2550 capacitance bridge) with respect to the initial distance between the lower sample mounting plates. Since the zero-volt distance of the capacitor plates is not independent of temperature (due to the thermal expansion of the epoxy holding these plates), first a calibration of the zero volt capacitance has to be determined as function of temperature. Ideally, this is done by affixing a relatively stiff bar of titanium (to match the thermal expansion of the cell made of titanium) across the two sides of the cell and measuring the capacitance as a function of temperature. Since the zero volt capacitance changes between each measurement run due to the hysteresis of the PZT stacks, it is important to record this value before the mounting procedure of each sample. The calibration can then be adjusted from measurement to measurement. The change of the capacitance with changing distance of the capacitor plates, on the other hand is almost temperature independent (the thermal contraction of the plates as well as the change of the dielectric permittivity of helium gas within the experimental temperature range are small). The manufacturer supplied calibration is thus almost temperature independent.

Due to strain relaxation effects within the glue layers and the Ti mounting plates, the actual strain on the sample differs from the nominal value. In this work, we estimated the strain relaxation effects using finite element simulations (more details given below). The strain experienced by the sample was calculated to be $\epsilon_{xx}$=($0.7\pm 0.07$)$\epsilon_{xx}^{disp}$. The difference in strain on the top and bottom surface of the sample was found to be less than 2\%, despite the asymmetric glue layer thicknesses. The normal strains within the sample were found to be approximately constant in distances over 100\,$\mu$m of the plates. The voltage contacts on our samples were placed so that only the section of the sample experiencing uniform strain was probed. To avoid shear strains introduced into the sample by asymmetric, point shaped contacts, the voltage (and the current) contacts were line shaped and spanned the entire width of the sample (see Fig.\,6b).

\MI{\subsubsection{Uniaxial stress along the tetragonal [110] axis}
\label{sec:ExpUniax110}
$B_{\rm 2g}$ strain is known to turn the phase transition into a cross over and smear all the related features. This is indeed what we observe in our experiment applying uniaxial stress along the tetragonal [110] axis. Figure\,\ref{fig:FigureSM4} shows the data for the $B_{\rm 2g}$ strain experiment. The feature corresponding to the nematic phase transition (step change in the derivative of the resistivity with respect to temperature) is quickly suppressed and replaced by a broad cross over. A relatively small strain $\epsilon_{xx}^{disp}$ of about $3\times10^{-3}$ is sufficient to fully suppress the feature associated with the nematic phase transition. Considering the evolution of the feature of the nematic phase transition in our $B_{\rm 1g}$ data presented in the manuscript, (and the rotated compliance tensor), we estimate a misalignment of 2.25$^\circ$. As the feature associated with the nematic phase transition stays sharp and clearly observable within the whole investigated strain range, the contamination (which to some extent might also originate from our voltage contacts) has no significance for our results.

\begin{figure}[ht]
	\centering
		\includegraphics[width=0.48\textwidth]{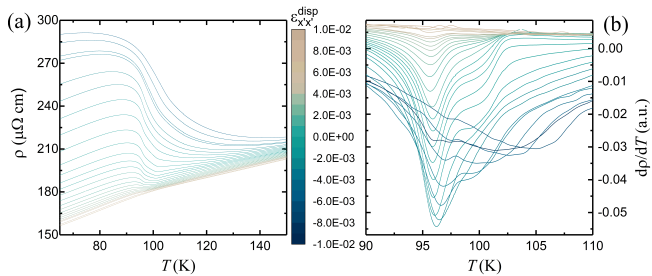}
		\caption{(a) Resistivity as a function of temperature for different strains $\epsilon_{\rm xx}^{disp}$, applied along the tetragonal [110] axis. (b) Derivative of the electrical resistivity with respect to temperature. The step like feature associated with the structural phase transition is suppressed quickly and is fully smeared for strains on the order of $\epsilon_{\rm xx}^{disp}\approx$$3\times10^{-3}$.}
	\label{fig:FigureSM4}
\end{figure}}

\subsection{Hydrostatic pressure experiments}
\label{sec:ExpHydro}
Hydrostatic pressure experiments were performed using a Quantum Design HPC-30 Cu-Be based self-clamping pressure cell. Although this version is no longer commercially available, information on the very similar updated version, HPC-33, can be found on the Quantum Design website. Hydrostatic pressure up to $\sim$3Gpa is applied using a hydraulic press. Daphne Oil 7373 is used as a pressure transfer medium. Note that the freezing point of the Daphne oil is always below room temperature for pressures less than 2\,GPa. This ensures a high degree of hydrostaticity throughout the experimental range \cite{Tor15.1}. Pressure measurements were performed by probing the superconducting transition temperature of a lead manometer \cite{Smi69.1}. In addition, the temperature dependence of the hydrostatic pressure within the HPC-30 pressure cell was determined by calibration measurements using both, a lead and a manganin manometer \cite{Tho84.1}. Below 100\,K the hydrostatic pressure was found to be almost independent of temperature.
 
\section{Finite element simulations}
\label{sec:Simul}
The goal of our finite element simulation was to estimate the strain relaxation effects in the glue layers as well as the Ti mounting plates.
The mechanical properties used within our simulations are summarized in Tab.\,\ref{tab:1} below.

\begin{table*}[ht!]
\caption{Summary of the mechanical properties used for our finite element simulations. For the tetragonal Ba(Fe$_{0.975}$Co$_{0.025}$)$_2$As$_2$, the Young's modulus is given for stress along [100] ($E$) and [001] ($E^{'}$). In addition, the table also shows the in-plane ($\nu$) and out-of-plane  Poisson ratio ($\nu^{'}$), as well as the in-plane ($C_{66}$) and out-of plane shear moduli ($C_{44}$). These properties correspond to the mechanical properties of Ba(Fe$_{0.963}$Co$_{0.0375}$)$_2$As$_2$ at 100\,K, as found by resonant ultrasound spectroscopy \cite{Yos12.1}. The mechanical properties of the mounting epoxy (Devcon 2-ton epoxy) at 100\,K were assumed to be slightly softer compared to filled Stycast 2850FT \cite{Hic14.1}.}
\begin{tabular}{ l c c c c c c}
  Material                                & $E$ (GPa)& $E^{'}$ (GPa)& $C_{66}$ (GPa)& $C_{44}$ (GPa)& $\nu$ & $\nu^{'}$\\
  \hline\hline			
  Titanium grade2                         & 105      & -          & 39.5     & -          &  0.33 &  - \\
  Epoxy                                   & 10       & -          & 3.8      & -          &  0.3  &  - \\
  Ba(Fe$_{0.975}$Co$_{0.025}$)$_2$As$_2$  & 82       & 82         & 10       & 39         &  0.26 &  0.164\\
  \hline  
\end{tabular}
\label{tab:1}
\end{table*}

The elastic properties of Ba(Fe$_{0.975}$Co$_{0.025}$)$_2$As$_2$ were estimated using the elastic stiffness tensor for 3.7$\%$ Co doped BaFe$_2$As$_2$ at 100\,K \cite{Yos12.1}. The mechanical properties of the epoxy were estimated to lie in between the properties of filled Stycast 2850FT and unfilled Stycast 1260 \cite{Hic14.1}. To our knowledge, the actual mechanical properties of Devcon two-ton epoxy are not characterized down to 100\,K. To justify the assumed mechanical properties we compared measurements on samples mounted using Devcon two-ton epoxy to measurements using Stycast 2850FT. Judged from the strain dependence of the structural and magnetic transition temperature of Ba(Fe$_{0.975}$Co$_{0.025}$)$_2$As$_2$, Devcon two-ton epoxy yields only a moderately smaller strain transmission as compared to Stycast 2850FT. Since the exact mechanical properties are unknown, we estimated the associated systematic error by varying the mechanical properties used within our finite element simulations by $\pm$50\%.
\begin{figure}[ht!]
	\centering
		\includegraphics[width=0.48\textwidth]{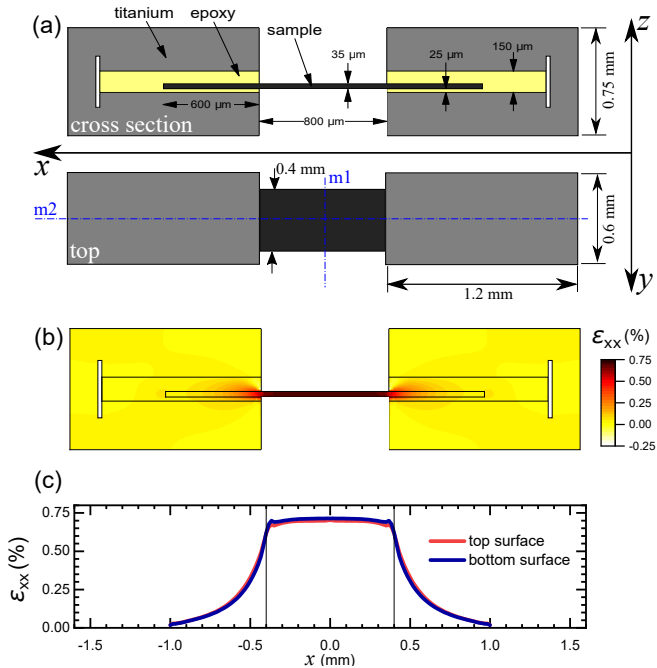}
	\caption{(a) Model used for the finite element simulations presented here. The dashed-dotted blue lines on the top view show the mirror planes $m_1$ and $m_2$ used to reduce the model size. (b) Results of our finite element simulation shown on the cross section of  our model. (c) Normal strain $\epsilon_{\rm xx}$ along the center line in $x$ direction on the top and bottom surface of the sample. The vertical black solid lines mark the position of the mounting plate edges. Nominally, a strain of $\epsilon_{\rm xx}$=1\% was applied. The strain relaxation within the mounting plates and the glue result in a strain transfer ratio of 0.7.}
	\label{fig:model}
\end{figure}

In order to minimize the computational requirements, the model shown in Fig.\,\ref{fig:model} was cut along mirror planes such that effectively only 25\% of the model had to be meshed and simulated. The model has been constrained such that the mounting plates were free to move only along a path parallel to $x$. A symmetric displacement corresponding to a reduction in the distance of the mounting plates by 8\,$\mu$m (corresponding to a nominal $\epsilon_{xx}^{\rm disp}$ of 1\%) was applied. For meshing, brick shaped elements were chosen. The element size was set to be 15\,$\mu$m. The contacts between the sample and the epoxy as well as the epoxy and the mounting plates were chosen to be perfectly rigid.

The strain within the sample along $x$ was found to be nearly constant 100\,$\mu$m away from the mounting plate edges. As can be seen from the results shown in Fig.\,9, out of plane shear strain (introduced by the asymmetric glue layer thickness on top and bottom) is small, resulting in a difference of $\epsilon_{xx}$ on the top and bottom surface of the sample of only about 2\% (Fig.\,\ref{fig:model}(c)). The strain transfer ratio $\epsilon_{xx}/\epsilon_{xx}^{\rm disp}$ used for analyzing our data was found to be $0.7 \pm 0.07$. The error bar was estimated from varying the mechanical properties of the epoxy within $\pm50$\% as well as the thickness of the bottom glue layer by $\pm$15$\mu$m.

\subsection{Elastic properties of Ba(Fe$_{0.975}$Co$_{0.025}$)$_2$As$_2$}
\label{sec:SimulElast}
Since we focus on vertical shear-free deformations (as indicated by our finite element simulations), the elastic constants $C_{\rm ij}$ with ${\rm  i,j} \leq 3$ (Voigt notation: $1 \equiv {\rm xx}, 2 \equiv {\rm yy} ,3 \equiv {\rm zz}$) fully describe the stress-strain relations relevant for our experiments. Out of the four independent elastic constants \footnote{$C_{\rm ij}=C_{\rm ji}$ due to the symmetry of the free energy derivatives $\frac{\partial}{\partial\epsilon_{\rm i}}\left(\frac{\partial F}{\partial\epsilon_{\rm j}}\right)=C_{\rm ij}$, and $C_{11}=C_{22}$ due to crystal symmetry}, $C_{11}$, $C_{12}$, and $C_{33}$ have recently been reported by a resonant ultrasound spectroscopy (RUS) study \cite{Yos12.1} for 3.7\% Co doped BaFe$_{2}$As$_2$ as a function of temperature. We here determine the missing $C_{13}$ by assuming an equal in- and out-of-plane Young's modulus (as suggested by a recent instrumented indentation experiment \cite{Des16.1}) and arrive at 
\begin{equation*}
\textbf{\textit{C}}=
\begin{pmatrix}
92.5 & 27.3 & 19.7 \\
27.3 & 92.5 & 19.7 \\
19.7 & 19.7 & 88.3 \\
\end{pmatrix}
{\rm GPa}
\end{equation*}
for 3.7\% Co doped BaFe$_{2}$As$_2$ at 100\,K. Using this tensor as an estimation for the shear-free mechanical properties of 2.5\% Co doped BaFe$_{2}$As$_2$ in the vicinity of its structural transition, it is straightforward to calculate the symmetry decomposed strain fields from the measured hydrostatic pressure and $\epsilon_{xx}$ for the two sets of experiments.

\section{Strain decomposition}
\label{sec:Decomp}
The relation between $\epsilon_{\rm A_{1g,1}}$=1/2$(\epsilon_{\rm xx}+\epsilon_{\rm yy})$ and $\epsilon_{\rm A_{1g,2}}$=$\epsilon_{\rm zz}$ under the two stress conditions studied here is determined by the elastic properties (the elastic stiffness tensor) of the investigated material. For hydrostatic pressure ($\sigma_{\rm xx}=\sigma_{\rm yy}=\sigma_{\rm zz}$), $\epsilon_{\rm A_{1g,2}}= \frac{1-2\nu'}{1-\nu-\nu'} \epsilon_{\rm A_{1g,1}}$, for uniaxial stress ($\sigma_{\rm xx}\neq0,\sigma_{\rm yy}=\sigma_{\rm zz}$=0), $\epsilon_{\rm A_{1g,2}}= \frac{-2\nu'}{1-\nu}\epsilon_{\rm A_{1g,1}}$, where $\nu$ and $\nu'$ are the in- and out-of-plane Poisson ratio.

Using these relations the total response of $T_{\rm S}$ to $\epsilon_{\rm A_{1g,1}}$ can be written as
\begin{equation}
\begin{split}
\left(\frac{{\rm d}T_{\rm S}}{{\rm d}\epsilon_{\rm A_{1g,1}}}\right)^{\rm hyd} &= \frac{\partial T_{\rm S}}{\partial \epsilon_{\rm A_{1g,1}}} + \frac{1-2\nu'}{1-\nu- \nu'} \frac{\partial T_{\rm S}}{\partial \epsilon_{\rm A_{1g,2}}}\\
\left(\frac{{\rm d}T_{\rm S}}{{\rm d}\epsilon_{\rm A_{1g,1}}}\right)^{\rm uni} &= \frac{\partial T_{\rm S}}{\partial \epsilon_{\rm A_{1g,1}}} - \frac{2\nu'}{1-\nu} \frac{\partial T_{\rm S}}{\partial \epsilon_{\rm A_{1g,2}}}
\end{split}
\label{eq1}
\end{equation}
for the hydrostatic pressure and the uniaxial stress experiment, respectively. Using the in- and out-of-plane Poisson ratio determined from the elastic stiffness tensor $\textbf{\textit{C}}$ ($\nu=0.26$ and $\nu'=0.164$) as well as the experimentally determined responses $\left(\frac{{\rm d}T_{\rm S}}{{\rm d}\epsilon_{\rm A_{1g,1}}}\right)^{\rm hyd}$ and $\left(\frac{{\rm d}T_{\rm S}}{{\rm d}\epsilon_{\rm A_{1g,1}}}\right)^{\rm uni}$, the above equations can be solved for $\frac{\partial T_{\rm S}}{\partial \epsilon_{\rm A_{1g,i}}}$. We find $\frac{\partial T_{\rm S}}{\partial \epsilon_{\rm A_{1g,1}}} = -6.35\frac{\rm K}{\%}$ and $\frac{\partial T_{\rm S}}{\partial \epsilon_{\rm A_{1g,2}}} = +16.7\frac{\rm K}{\%}$ for the investigated Ba(Fe$_{0.975}$Co$_{0.025}$)$_2$As$_2$ confirming the largest linear contribution comes from $\epsilon_{\rm A_{1g,2}}$. 

Errors reported in the paper  represent statistical uncertainty. A full error analysis considering also systematic uncertainty yields, $\frac{\partial{T_{\rm S}}}{\partial\epsilon_{\rm A_{1g},1}}=-6.35 \pm \left(0.23\frac{\rm K}{\%}\right)^{\rm statistical} \pm \left(2.28\frac{\rm K}{\%}\right)^{\rm systematic}$ and $\frac{\partial{T_{\rm S}}}{\partial\epsilon_{\rm A_{1g},2}}=16.7 \pm \left(0.32\frac{\rm K}{\%}\right)^{\rm statistical} \pm \left(1.45\frac{\rm K}{\%}\right)^{\rm systematic}$. The main systematic errors considered are a 5\% uncertainty in the elastic constants as well as the measured hydrostatic pressure, and a 10\% uncertainty in the strain relaxation factor. All errors were assumed to be uncorrelated. As the sign of our measured responses is robust within the error bars, our main results and discussion are unaffected by the systematic errors mentioned here.

\end{document}